\documentclass[aps,showpacs,epsf,twocolumn,prb]{revtex4}

\usepackage{amsmath}
\usepackage{amssymb}
\usepackage{bbold}
\usepackage{mathrsfs}
\usepackage{graphicx, psfrag}
\usepackage[centerlast]{caption}
\usepackage{float}
\usepackage{subfig}
\usepackage{tikz}
\usepackage{pgfplots}
\usepackage[colorlinks=true, citecolor=blue, urlcolor = blue, linkcolor= red, bookmarks=true]{hyperref}
\usepackage{epstopdf}

\begin{document}
\def \beq{\begin{equation}}
\def \eeq{\end{equation}}
\def \bse{\begin{subequations}}
\def \ese{\end{subequations}}
\def \bea{\begin{eqnarray}}
\def \eea{\end{eqnarray}}
\def \bem{\begin{displaymath}}
\def \eem{\end{displaymath}}
\def \bem{\begin{pmatrix}}
\def \eem{\end{pmatrix}}
\def \bc{\begin{center}}
\def \ec{\end{center}}
\def \bb{\bibitem}
\def \bs{\boldsymbol}
\def \nn{\nonumber}
\def \mf{\tilde{J}_1}
\def \mj{\tilde{J}_0}
\def \mh{\mathcal{H}}
\def \ma{\mathcal{A}}
\def \md{\mathcal{D}}
\def \mg{\mathcal{G}}
\def \hkx{\hat{k}_{x}}
\def \hky{\hat{k}_{y}}
\def \bq{\bar{q_{y}}}
\newcommand{\cc}{\color{red}}
\newcommand{\cb}{\color{blue}}
\newcommand{\upa}{\uparrow}
\newcommand{\dna}{\downarrow}
\newcommand{\pdag}{\phantom\dagger}
\newcommand{\bra}[1]{\langle #1 |}
\newcommand{\ket}[1]{| #1 \rangle}
\newcommand{\braket}[2]{\langle #1 | #2 \rangle}
\newcommand{\ketbra}[2]{| #1 \rangle \langle #2 |}
\newcommand{\expect}[1]{\langle #1 \rangle}
\newcommand{\inx}[1]{\int d\bs x \Big [ #1 \Big ]}
\newcommand{\si}[1]{\Psi^{\pdag}_{#1} (\bs x)}
\newcommand{\dsi}[1]{\Psi^\dag_{#1} (\bs x)}
\newcommand{\evolve}[1]{\frac{\partial #1}{\partial t}}

\title{Unconventional band structure for a  periodically gated surface of a three dimensional Topological Insulator}
\author{ Puja Mondal and Sankalpa Ghosh}
\affiliation{Department of Physics, Indian Institute of Technology Delhi, New Delhi-110016, India}

\begin{abstract}
The surface states of the three dimensional (3D) Topological Insulators are described by  two-dimensional (2D) massless dirac equation. A  gate voltage induced one dimensional potential barrier on such surface creates a discrete bound state in the forbidden region outside the dirac cone. Even for a single barrier it is shown such bound state can create  electrostatic analogue of Shubnikov de Haas oscillation which can be experimentally observed for relatively smaller size samples.  However 
when these surface states are exposed to a periodic arrangement of such gate voltage induced potential barriers, the band structure of the same  got nontrivially modified. This is expected to significantly alters the properties of macroscopic system. We also suggest that in suitable limit the system may offer ways to control electron spin electrostatically which may be practically useful. 
\end{abstract}
\maketitle
The discovery of two-dimensional quantum spin Hall insulator commonly known as two dimensional 
topological insulators (2DTI) \cite{KaneMele, Bernevig, Molenkamp} and the subsequent discovery of their three dimensional generalization dubbed as three dimensional  topological insulators (3DTI) \cite{Moore, Kane, Hasan1} 
led to a large amount of experimental and theoretical work in this direction \cite{Hasan, ZhangR}. 
The surface electronic states of the 3DTI are described by the two dimensional massless dirac equation and this has been demonstrated by spin and angle resolved photoemission spectroscopy \cite{Hasan1, Hasan2}.  Such massless dirac fermions (MDF) with ultra relativistic dispersion relation have fundamentally different transport properties from in comparison to the non-relativistic electron gas (NREG)  in ordinary metal or semiconductor. 

One such peculiar properties of these surface MDF is the formation of bound states in a one dimensional potential barrier \cite{Quantumdot, Nagaosa} created through a gate voltage outside 
the dirac cone, namely in the forbidden region. 
This situation should be contrasted with the prototype bound state and quasi-bound states formation in presence of quantum well in non-relativistic quantum mechanics ( for example see \cite{NRbound}) as well as for  the case of MDF in Graphene \cite{MDFbound1, MDFbound2, MDFbound3}. Particularly in the later case ( for example see \cite{MDFbound3})
these bound states formed by the quantum wells are  within the dirac cone which are in proximity with scattering states having linear dispersion. 

In this paper, we report such bound state induced significant modification of band structure for  surface MDF in presence of a periodic array of such barriers. 
This modification of band structure occurs outside the dirac cone which is otherwise a forbidden zone and in a nontrivial manner changes the band structure of such surface states. 
We start by showing  in presence of such bound states in a potential barrier  
the DOS of MDF in the surface of a 3DTI oscillates purely through electrostatic means creating electrostatic analogue of Shubnikov de Haas (SdH)oscillation of NREG in a magnetic field \cite{Solyom}. Such DOS oscillation leads to sharp oscillation in the conductance in the linear response regime. However, since the DOS scales with the relative size of the gated barrier region for macroscopic 
sample such conductance oscillation is hard to observe. To observe the effect of such bound states in the macroscopic sample we therefore consider a periodic array of such barriers on the surface of 3DTI in this paper.  We show that the resulting band structure is unique for such  MDF and 
consists of two distinct part, one inside the dirac cone formed out of continuum scattering states 
and the other outside the dirac cone originating from the bound states. When we consider the potential barrier in the $\delta$-function limit, the corresponding bound states are one dimensional helical states with spin and momentum locked. In a periodic set-up of such $\delta$-function potentials bands formed by 
helical states may provide methods to control spin through electrostatic means. 

 \begin{figure}
\subfloat{\includegraphics[width=0.7\columnwidth,height=0.3 
\columnwidth]{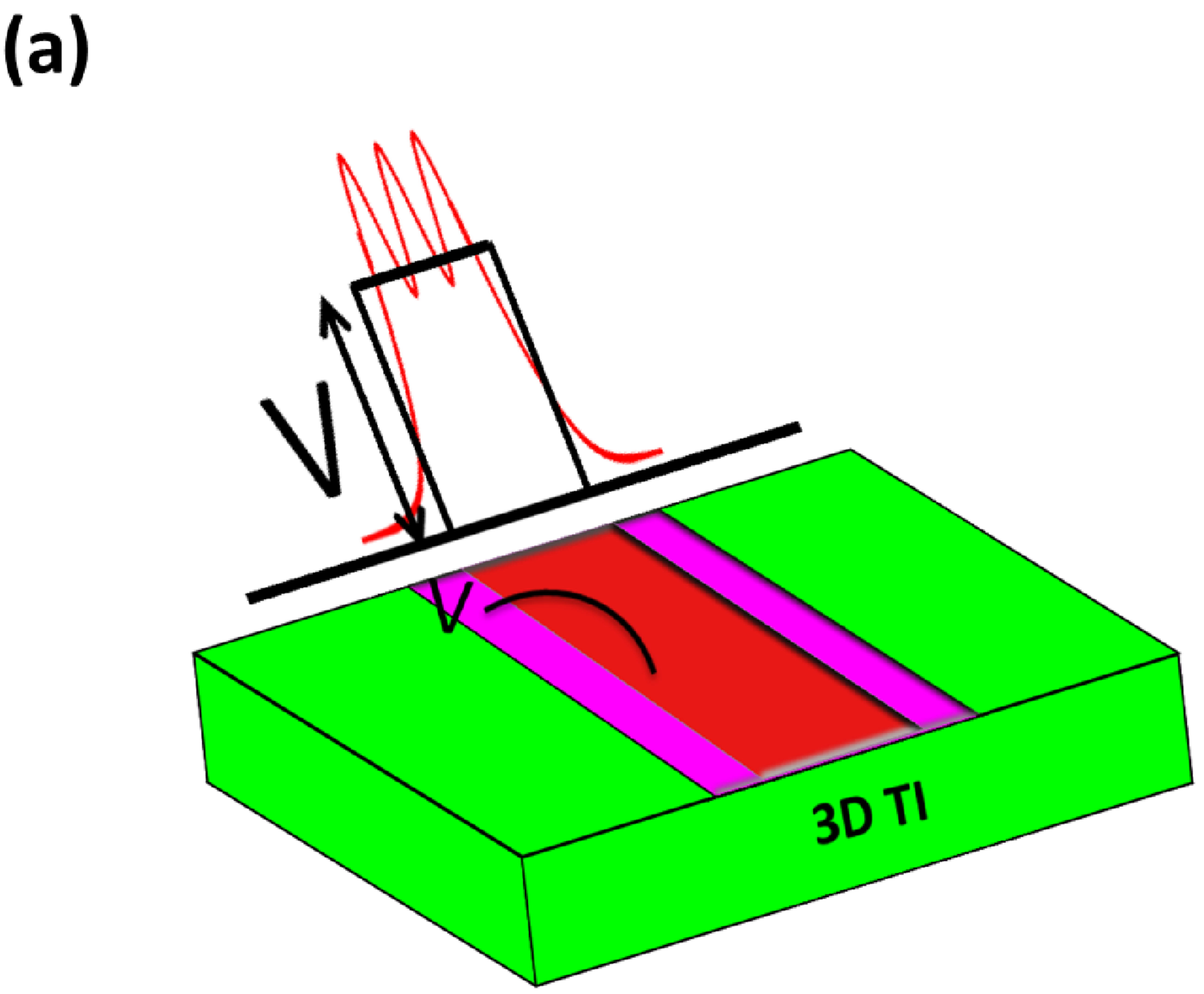}} \\
\subfloat{\includegraphics[width=0.7\columnwidth,height=0.3
\columnwidth]{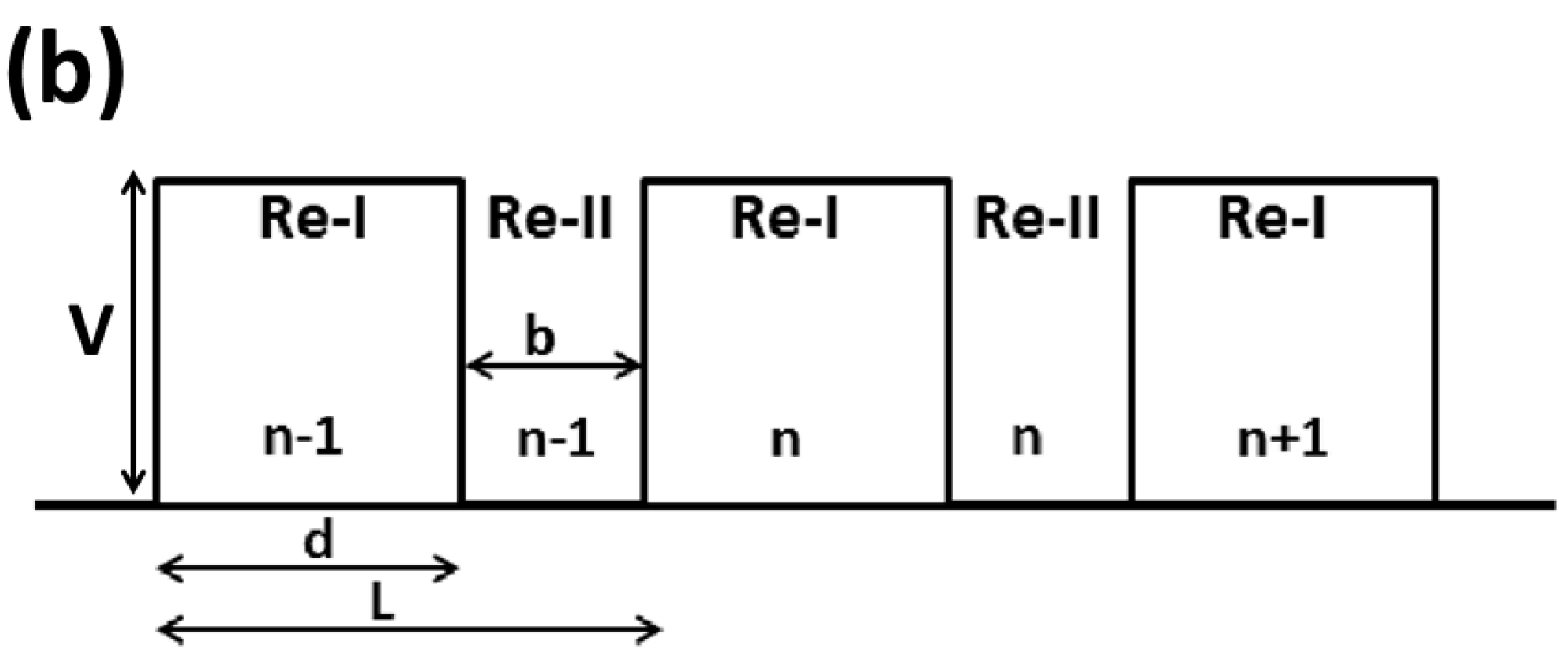}}
\caption{{\it (color online)} (a) Schematic of the potential barrier and the bound state wavefunction(red) on the surface of a 3DTI (b) Schematic figure of 
periodic arrangement of such potential barrier. Here $V=V_{0}$ in both figures.}
\label{fig:periodic} 
\end{figure}
The effective hamiltonian describing surface states of 3DTI can be written as 
\beq
H_{tot}=v_{F}(\bs{\sigma}~.~\bs{p})+\frac{\lambda}{2}\sigma_{z}(k_{+}^{3}+k_{-}^{3})
\eeq
Here $v_{F}$ and $\lambda$ are the Fermi velocity and wrapping parameter. $\bs{\sigma}=\sigma_{x} \hat{i} + \sigma_{y} \hat{j}$ is the Pauli matrix vector describing the real spin of the electron and $k_{\pm}=k_{x}\pm ik_{y}$. The first term in the hamiltonian corresponds to  that for two dimensional MDF  giving circular shaped energy contour for ungated 3DTI surface states centered artound the Dirac point and dominates upto certain energy value (eg. in the case of $Bi_{2}Te_{3}$ it is up to  $150~meV$ and in case of $Bi_{2}Se_{3}$ it is up to 100 meV).
The contribution to the Hamiltonian due to hexagonal wrapping (HW) effect is given by the second term $\frac{\lambda}{2}\sigma_{z}(k_{+}^{3}+k_{-}^{3})$ which becomes more effective as one move away from the Dirac point \cite{Fu,Kuroda} and shows deviation from the circular energy plot. Therefore in the vicinity of the Dirac point  within the above mentioned energy range  the surface states of 3DTI have same Dirac like Hamiltonian as Graphene, namely  
\begin{equation}
H=v_{F}(\bs{\sigma}~.~\bs{p}) \label{diracham}
\end{equation}  
but without valley degeneracy like the later.

The other approximation that is included in the hamiltonian \ref{diracham}  is that the anisotropy in the Fermi velocity is ignored \cite{Fu}. This is again valid in the close vicinity of Dirac point. We also consider here a single Fermi corssing for the surface states as opposed to  more number(odd number higher than one) of fermi crossing of the surface states \cite{Hsieh}. Thus we model the surface states of the 3DTI with the the hamiltonian (\ref{diracham}) assuming that the height of the potential barrier (\ref{potbarrier}) is within the stipulated limit satisfying the above mentioned conditions.

It may be also noted that such surface states can alternatively be described by the effective hamiltonian $H'= (v_{f}\bs{\sigma} \times \bs{k})_{z}$ \cite{Fu} which can be obtained from (\ref{diracham}) through a unitary transformation. 
We consider such surface states in a scalar potential barrier (Fig. \ref{fig:periodic} (a))
\begin{equation}
 V(x) =
  \begin{cases}
  V_{0} & \text{if }\mid x\mid < d/2 \\
  0       & \text{if } \mid x\mid > d/2
  \end{cases}
   \\
\label{potbarrier}\end{equation}
which only varies along the $x$-direction. We chose the height of the potential barrier should be less than the bulk gap of a 3DTI so that it does not create bulk excitation in the system. As known from the experimental work, 
the 3DTI has lagre  bulk band gap of the order of 0.3 eV for $Bi_{2}Se_{3}$ \cite{Xia} and 0.15 eV for $Bi_{2}Te_{3}$\cite{Chen}. It may be also noted that such potential respects the time reversal symmetry of these surface states. 

Several comments are in order to justify the use of the effective hamiltonian (\ref{diracham}) to model the surface states of a 3DTI  and  to 
decide about the typical value of the potential barrier (\ref{potbarrier}) for which the effect described in the current work can be observed for a realistic 3DTI 
surface. The model hamiltonian (\ref{diracham}) describes massless dirac fermions with zero chemical potential which is strictly valid only at (or in the immediate neighborhood of ) 
the dirac point.\\

Writing the stationary solutions of the Schr\"odinger equation with energy $E$ as  
 $\psi(x,y)=\psi(x)~e^{iq_{y}y}~e^{-iEt}$
for a given $V_{0}$,  $\epsilon = \frac{E}{\hbar v_{F}}$,  
the $x$-component of the wave vector is given by 
\bea q_{x} & =  & \sqrt{\epsilon^{2} - q_{y}^{2}}, |x| \ge \frac{d}{2} \nonumber \\
                 & = & \sqrt{ ( \frac{E-V_{0}}{\hbar v_{F}})^{2} - q_{y}^{2}}, |x| < \frac{d}{2} \label{qx} \eea 

\begin{figure}
\subfloat{\includegraphics[width=0.45\columnwidth,height=0.4 
\columnwidth]{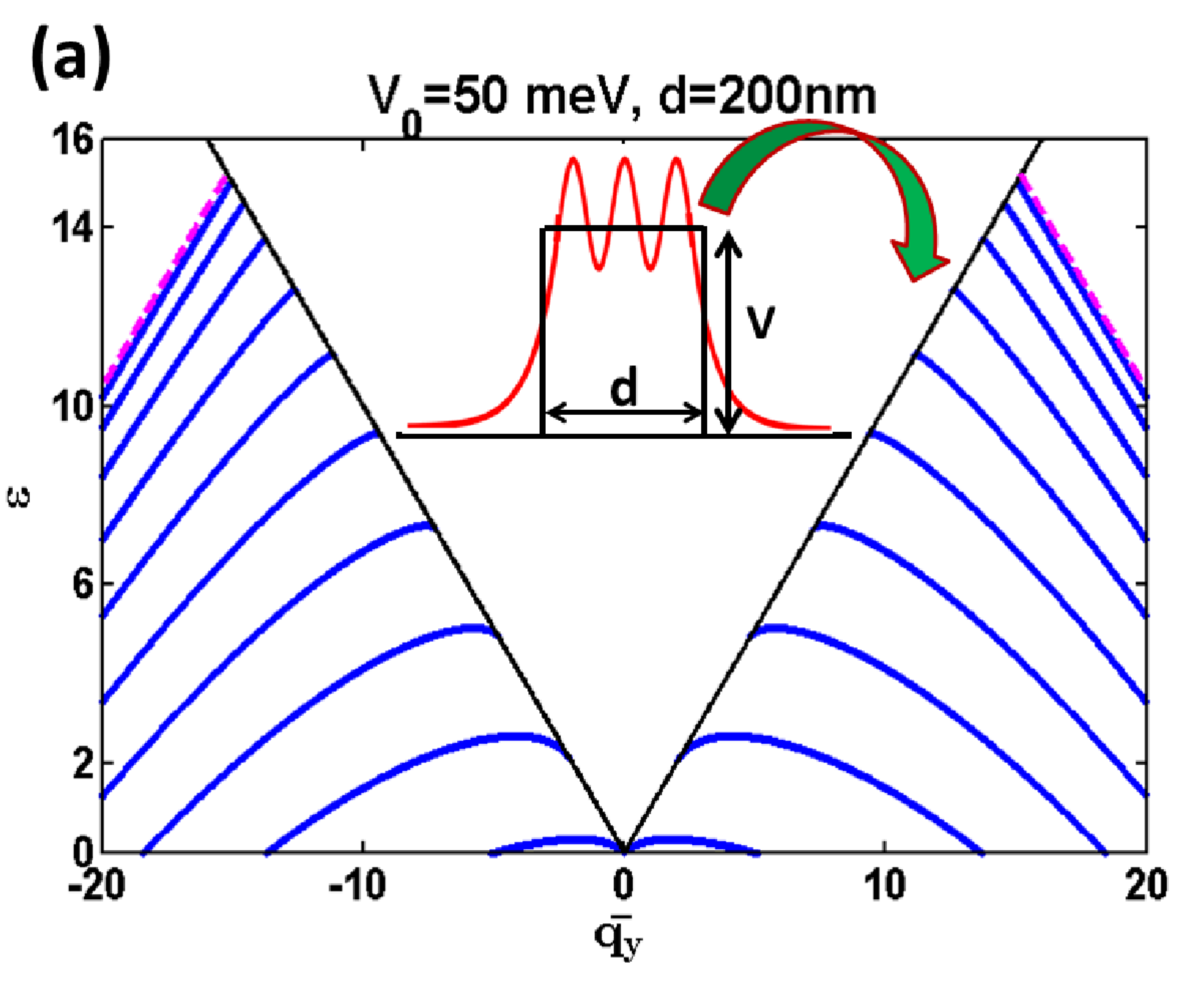} }
\subfloat{\includegraphics[width=0.45\columnwidth,height=0.4
\columnwidth]{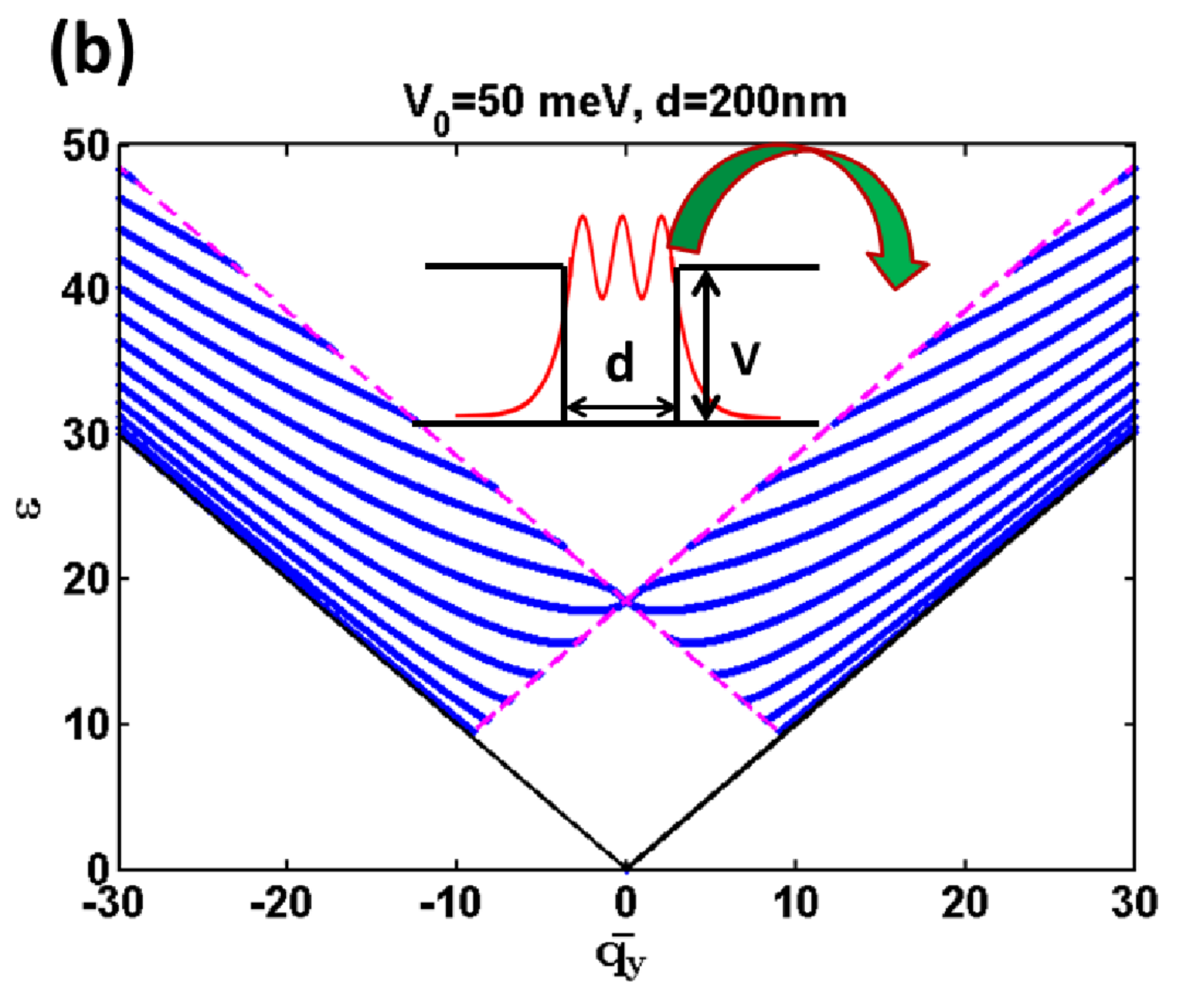} } \\
\subfloat{\includegraphics[width=0.45\columnwidth,height=0.4
\columnwidth]{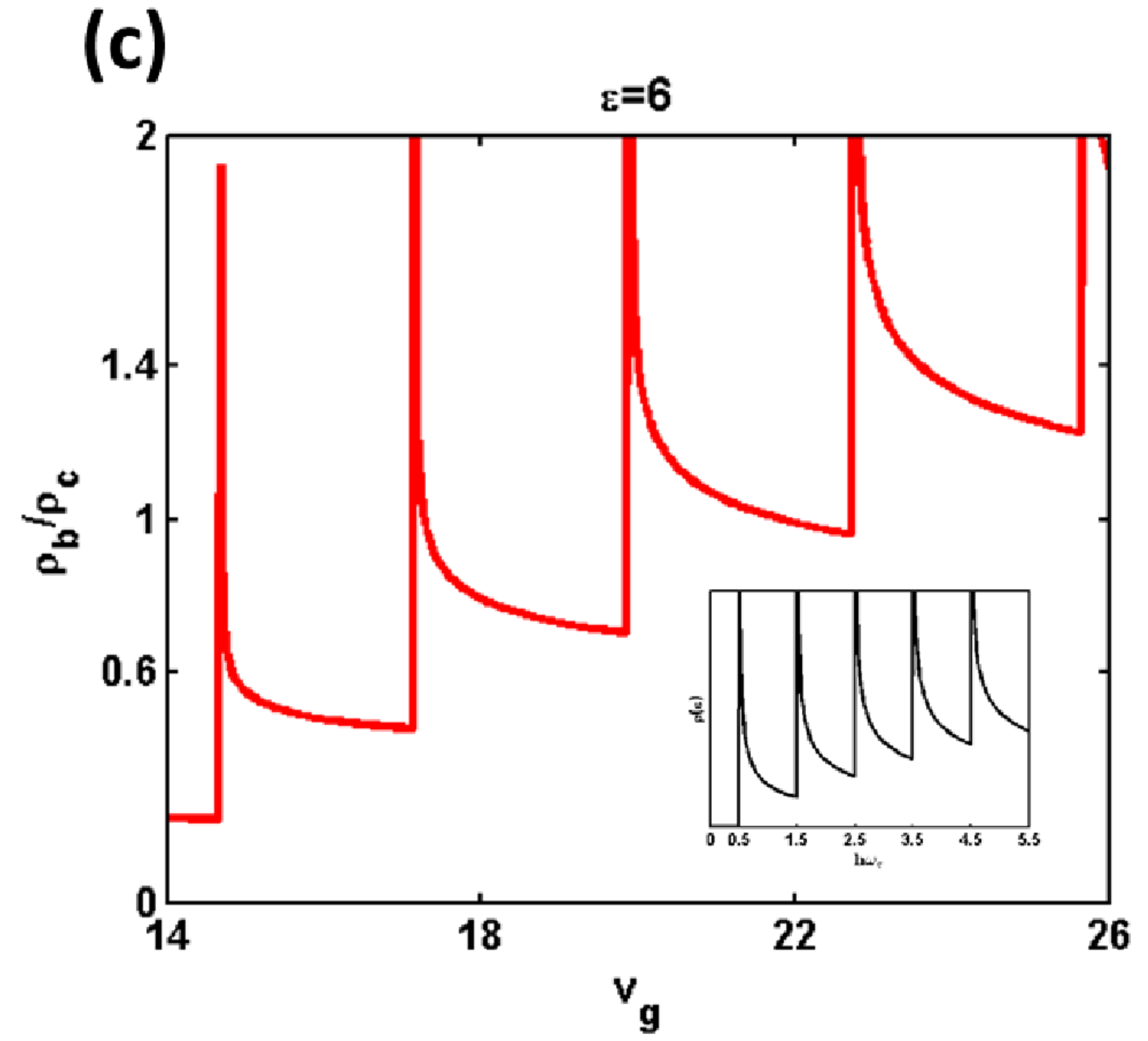}}
\subfloat{\includegraphics[width=0.45\columnwidth,height=0.4
\columnwidth]{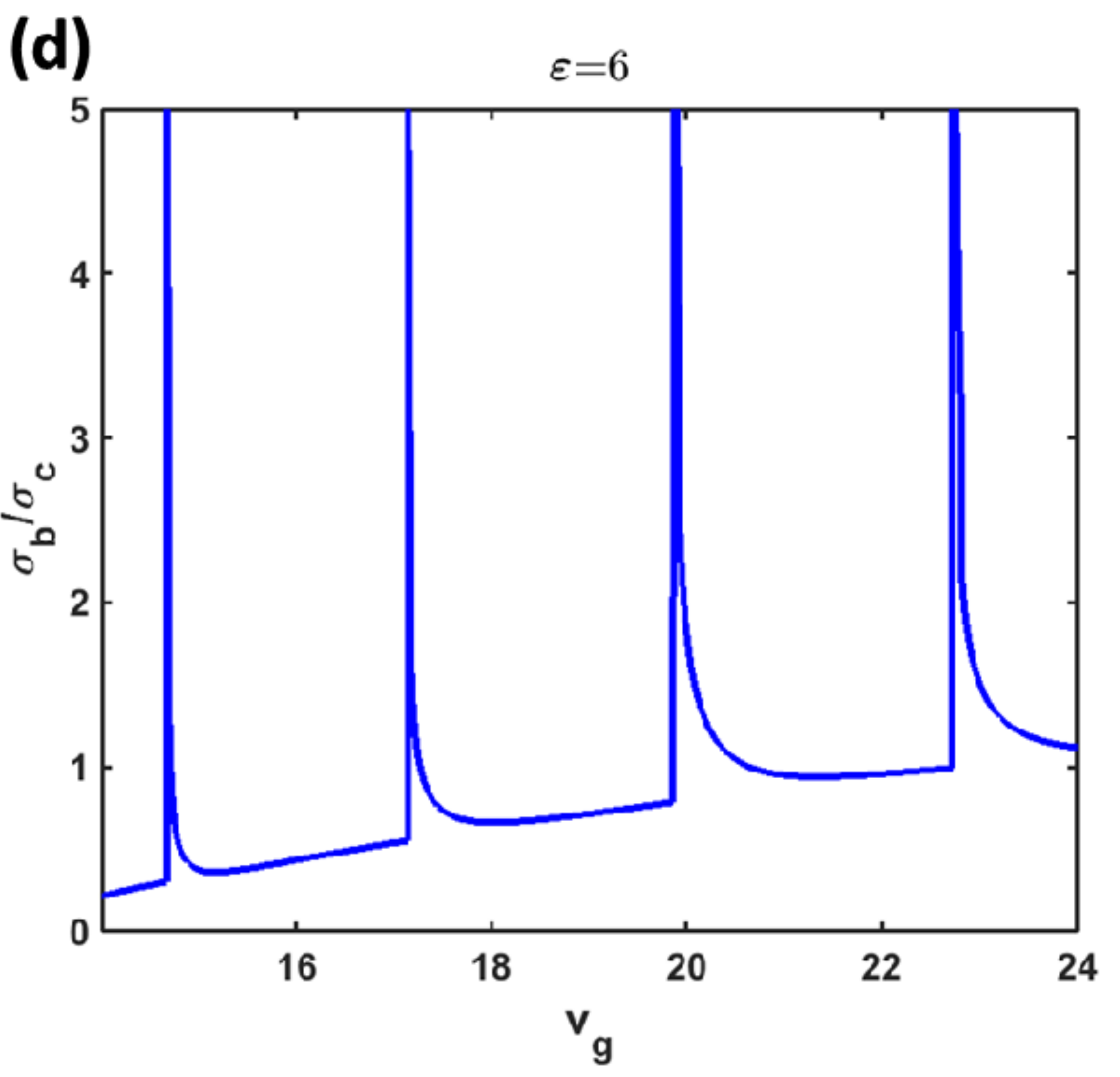}}
\caption{(Color online) (a) Bound states (blue) for potential barrier outside the dirac cone ( black lines) in between $\varepsilon= \pm\vert \bq \vert $ (black) to $\varepsilon=v_{g} \pm \vert \bq \vert$ (magenta)
for a potential barrier 
(b) For comparison bound states inside the dirac cone ( black line) for the potential well problem for MDF ( see the discussion in \cite{Suppli} sec.I ) are plotted. 
(c) relative DOS due to bound states 
as a function of $v_{g}$. In the inset similar DOS for a a NREG in a uniform magnetic field $B\hat{z}$ ($\omega_{c}=\frac{eB}{mc}$) is shown (d) Conductance oscillation due to bound states as a function of $v_{g}$}
\label{fig:DOS} 
\end{figure}

We define $\kappa=iq_{x}$, $\alpha = \tanh ^{-1} (\frac{\kappa}{q_{y}})$,  
$q_{x}=\sqrt{-q_{y}^{2}+(v_{g}/d-\epsilon)^{2}},~~ v_{g}=V_{0}d/\hbar v_{F}$ as the effective barrier strength,   $\theta=tan^{-1}(\frac{q_{y}}{q_{x}})$.
Eq. (\ref{qx}) shows that apart from the usual scattering solutions with 
$\epsilon >|q_{y}|$, there exist bound state solutions in otherwise forbidden zone $\epsilon <|q_{y}|$. For such solutions the $x$-component of the wave vector is imaginary outside the barrier regime, whereas it is real inside the barrier region and such solutions exist if 
$\epsilon< \vert q_{y} \vert < \vert v_{g}/d-\epsilon \vert  $. This condition can only be staisfied with linear dispersion for the MDF.
Such type of bound states can not be created in case of 2D NREG with quadratic dispersion.  
The wavefunctions  for such solutions are 
\bea
\psi(x) & = & 
  \begin{cases}
  &Ae^{\kappa (x+d/2)}\left( \begin{array}{cc}
  1\\
   \exp ( i ( \frac{\pi}{2} + i \alpha)) 
\end{array} \right),~~ x<-d/2\\
  & Be^{-\kappa(x-d/2)}\left( \begin{array}{cc}
  1\\
   \exp ( i (\frac{\pi}{2} - i \alpha))
\end{array} \right),~~x>d/2 
  \end{cases}
\label{bstateout} \\
\psi(x) & =&  Ce^{iq_{x}x}\left( \begin{array}{cc}
  1\\
  e^{i\theta}
\end{array} \right)+D e^{-iq_{x}x}\left( \begin{array}{cc}
  1\\
  -e^{-i\theta}
\end{array} \right),\mid x\mid < d/2~~\nn\\
\label{bstatein}
\eea
A schematic profile of such bound state wave function is given in Fig.\ref{fig:periodic}(a).  
The continuity of the  wave function at $x=\pm d/2$ determines $A,B,C,D$ whose nontrivial solutions 
 gives the quantization condition  
%
\begin{equation}
\textit{tan} ~\sqrt{(\varepsilon -v_{g})^2-\bq^2}+\dfrac{\sqrt{\bq^2-\varepsilon^2} \sqrt{(\varepsilon -v_{g})^2-\bq^2}}{\bq^2+\varepsilon(v_{g}-\varepsilon)}=0.
\label{bound} \end{equation}
Here $\bq=q_{y}d$ and $\varepsilon=\epsilon d$ are dimensionless. Eq.(\ref{bound}) can be solved numerically to yield the bound states solutions, 
bounded between two parallel lines from $\varepsilon= \pm\vert \bq \vert $ to $\varepsilon=v_{g} \pm \vert \bq \vert$ (Fig.\ref{fig:DOS} (a)).
These bound state solutions have real energy and exists outside the dirac cone. 
The situation is contrasted with the bound state formation for massless dirac fermions inside a potential well ( see Fig. \ref{fig:DOS}(b)) ( for details see \cite{Suppli}).  
The quantized energy values for a given value of 
the gate voltage is given by $\varepsilon_{n}  = \dfrac{v_{g}}{2}-\dfrac{n^{2}\pi^{2}}{2v_{g}}.$

Such gate voltage tunable discrete number of bound states in the energy spectrum profoundly effects the DOS and consequently other properties. Here  $v_{F}=5\times10^{5}$ m/s of $Bi_{2}Si_{3}$ for our calculation \cite{Zhang}.
For the continuum states of MDF on the surface of a 3DTI obeying  $\epsilon(\bs{q})= \mid \bs{q} \mid$, DOS  $\rho_{c}$ is 
\beq \rho_{c}(\epsilon) = \dfrac{2L_{x}L_{y}}{4\pi^2}\int d^{2}q \delta(E-E(q)) = \rho_{0}|\varepsilon|\label{DOSC} \eeq
with $\rho_{0}=\frac{L_{x}L_{y}}{\pi d \hbar v_{F}}$, $L_{x,y}$ is the sample length along $x,y$. 
The contribution to the DOS due to the discrete bound  states ($\rho_{b}$) can be calculated from 
Eq. (\ref{bound}) as 
\bea
\rho_{b}(\varepsilon) &= & 2\rho_{0} \dfrac{d}{L_{x}}\sum_{n} \mid \dfrac{d\bq}{d\varepsilon_{n}} \mid_{\varepsilon_{n}(\bq)=\varepsilon}
\label{DOSB} \\
\text{where}~ 
\dfrac{d\bq}{d\varepsilon} & = & \bq\dfrac{v_{g}-2 \varepsilon+(v_{g}-\varepsilon)\sqrt{\bq^2-\varepsilon^2}}{\varepsilon(v_{g}-\varepsilon)-\bq^2-\bq^2\sqrt{\bq^2-\varepsilon^2}}
\eea 
 $\rho_{b}$ scales with $d$ and  its variation with $v_{g}$ is plotted in Fig.\ref{fig:DOS}(c). When for a given energy the condition $\varepsilon  = \varepsilon_{n}$ 
is satisfied for a given $v_{g}$,  a jump occurs in DOS as expected from Eq. (\ref{DOSB}). 
From Fig.\ref{fig:DOS} (c) one finds that the behavior of $\rho_{b}$ as a function of the gate voltage is very similar to that of the DOS of a NREG in presence of a magnetic field. To show how such DOS influences the transport, we calculate the conductance in presence of  such bound states.

Since the DOS receives contribution both from the free massless dirac fermions as well as the bound states, either of these 
states contribute to the conductivity tensor. The conductance of free 2D MDF was already studied \cite{Ludwig, Shankar, DStone}. 
Briefly, in terms of  energy eigen states the expression
for the frequency ($\omega$) dependent conductivity tensor at finite temperature ($T$)
\bea \sigma_{\mu \nu} ( \omega, \beta) & = & \frac{i}{\omega} \int d \varepsilon \int \frac{d \bs{r} d \bs{r}'}{L_{x}L_{y}} 
\sum_{m,n} \langle m | \hat{j}_{\mu}(\bs{r}) | n \rangle \langle \hat{j}_{v}(\bs{r}') | m \rangle \nonumber \\
&  & \delta ( \varepsilon - \varepsilon_{m} ) \frac{ f (\varepsilon) - f(\varepsilon_{n})}
{\varepsilon - \varepsilon_{n} + \hbar (\omega + i \delta')} \label{Kubo} \eea 
where $f(\varepsilon) = \frac{1}{\exp(\beta \varepsilon) +1}$ is the Fermi-Dirac distribution at temperature $T=k_{B}(\beta)^{-1}$ and 
zero chemical potential. By taking $L_{x,y} \rightarrow \infty$, then $\delta' \rightarrow 0 $ and $\omega \rightarrow 0$ in (\ref{Kubo})
expression for d.c. conductivity can be obtained.  

In this paper our main purpose is to see change in the conductivity due to presence of bound states. Therefore the contribution to the conductivity by the scattering electrons is used as a overall scale factor for such boundate induced conductivity calculated upto the leading order. To this purpose 
we have only considered the first term in Kubo formula in the linear response 
regime to calculate conductivity for continuum and bound states without considering any vertex correction. Scattering of surface electrons are intrinsically anisotropic because of the fact that surface states $\vert \psi (\vec{k})\rangle$ and $\vert \psi (-\vec{k})\rangle$ form Kramers pair and they are orthogonal. In this case transport time of surface electrons is not equal to scattering time of ordinary electron in presence of disorder. Transport time is equal to the twice of the scattering time of conventional electrons for scalar isotropic disorder i.e $\tau_{tr}=2\tau_{e}$. 
 
 The next leading correction to the conductivity calculated here is the  vertex correction. This will add another term to the classical conductivity which will be from correction due to ladder diagram or diffusion. In this case re normalized vertex current is proportional to the bare current. The contribution of diffusion to the conductivity will be the same order of bare current. Corresponding calculation for surface states of 3DTI was performed in literature \cite{Carpentier}. 
Contribution to the conductivity from diffusion is of the form 
\beq
\sigma_{x} = \frac{\hbar}{2\pi}tr \left[ J_{x} \Gamma^{(d)} J_{x} \right] \nonumber
\eeq
where $\Gamma^{(d)}$ and $J_{x}$ are diffusion structure factor and re normalized vertex current. Here we have not included such contributions. 
 Within the above mentioned apprximations the expression of zero temperature d.c. conductance for MDF due to the continuum of the scattering states is obtained as ( details in \cite{Suppli})
\beq \sigma_{yy}^{c} (\omega \rightarrow 0) = 
\dfrac{e^{2} L_{x}\pi}{ L_{y}\hbar}[\dfrac{\epsilon_{F}\tau_{tr}}{\hbar}+\dfrac{1}{\pi}(1-\dfrac{\epsilon_{F}\tau_{tr}}{\hbar}\tan^{-1}(\hbar/\epsilon_{F}\tau_{tr}))]
\eeq
where $\tau_{tr}=2\tau_{e}$ is the transport time of surface electron in presence of disorder. 
Similarly the expression of the conductance in the $\omega \rightarrow 0$ and $ T \rightarrow 0$ limit due to the bound states can be calculated as (details in \cite{Suppli})
\beq \sigma_{yy}^{b} = \dfrac{4 e^{2}d} { \pi L_{y}\hbar}\sum_{n} \vert \dfrac{d\bq}{d\varepsilon}\vert_{\varepsilon_{n}(\bq)=\varepsilon_{F}}~\frac{\chi(\varepsilon_{F})}{\eta}
\label{conloc} \eeq 
where  $\chi_{n}(\varepsilon)=\vert\int_{x} \psi_{n}^{ q_{y} \dagger}(x)\sigma_{y}\psi_{n}^{q_{y}}(x) \vert^{2}$.
The ratio  of the free particle and bound state contribution to $\sigma_{yy}$ (we drop common $_{yy}$),
\beq
\dfrac{\sigma^{b}}{\sigma^{c}}=\dfrac{4 d}{\pi L_{x}\varepsilon_{F}}\vert \dfrac{d \bq}{d\varepsilon}\vert_{\varepsilon_{n}(\bq)=\varepsilon_{F}}\chi(\varepsilon_{F})
\label{condosci} \eeq
which oscillates with the changing $v_{g}$. This is plotted in Fig. \ref{fig:DOS} (d).  As expected 
this oscillation is similar to the SdH oscillation in presence of magnetic field due to the the discrete nature of the bound states. As the current oscillation  suggest, a gated surface of the 3DTI can therefore be used for switching purpose \cite{switch}. 
However, unlike in the case of SdH oscillation, here $\sigma_{b}$ scales with the 
$\frac{d}{L_{x}}$, the relative width of the barrier. Therefore whereas for a mesoscopic sized sample, such single barrier induced oscillation 
may be observed \cite{Nori}, in a macroscopic sample such effect will vanish. 

\begin{figure}
\subfloat{\includegraphics[width=0.5\columnwidth,height=0.4 
\columnwidth]{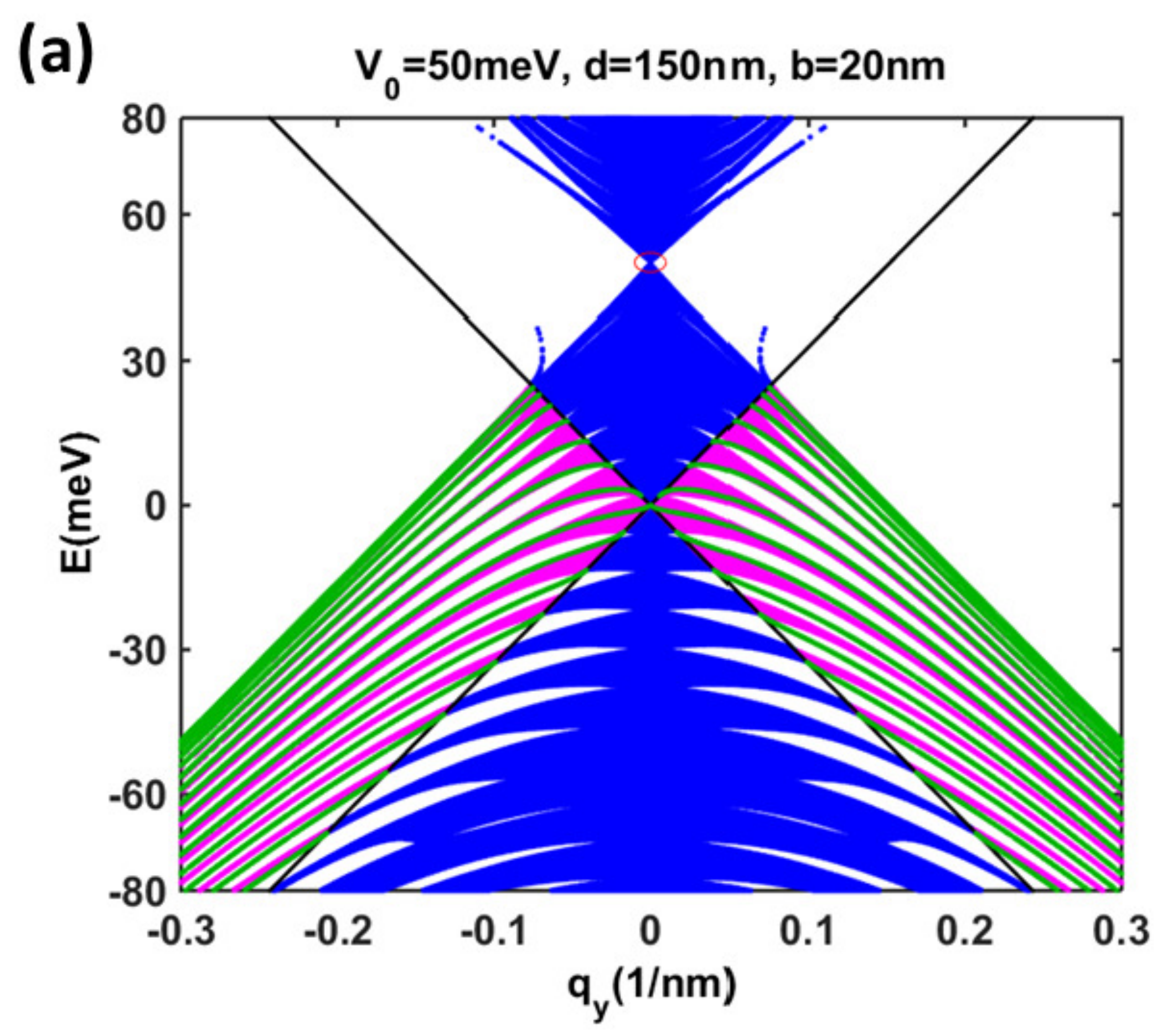} }
\subfloat{\includegraphics[width=0.5\columnwidth,height=0.4
\columnwidth]{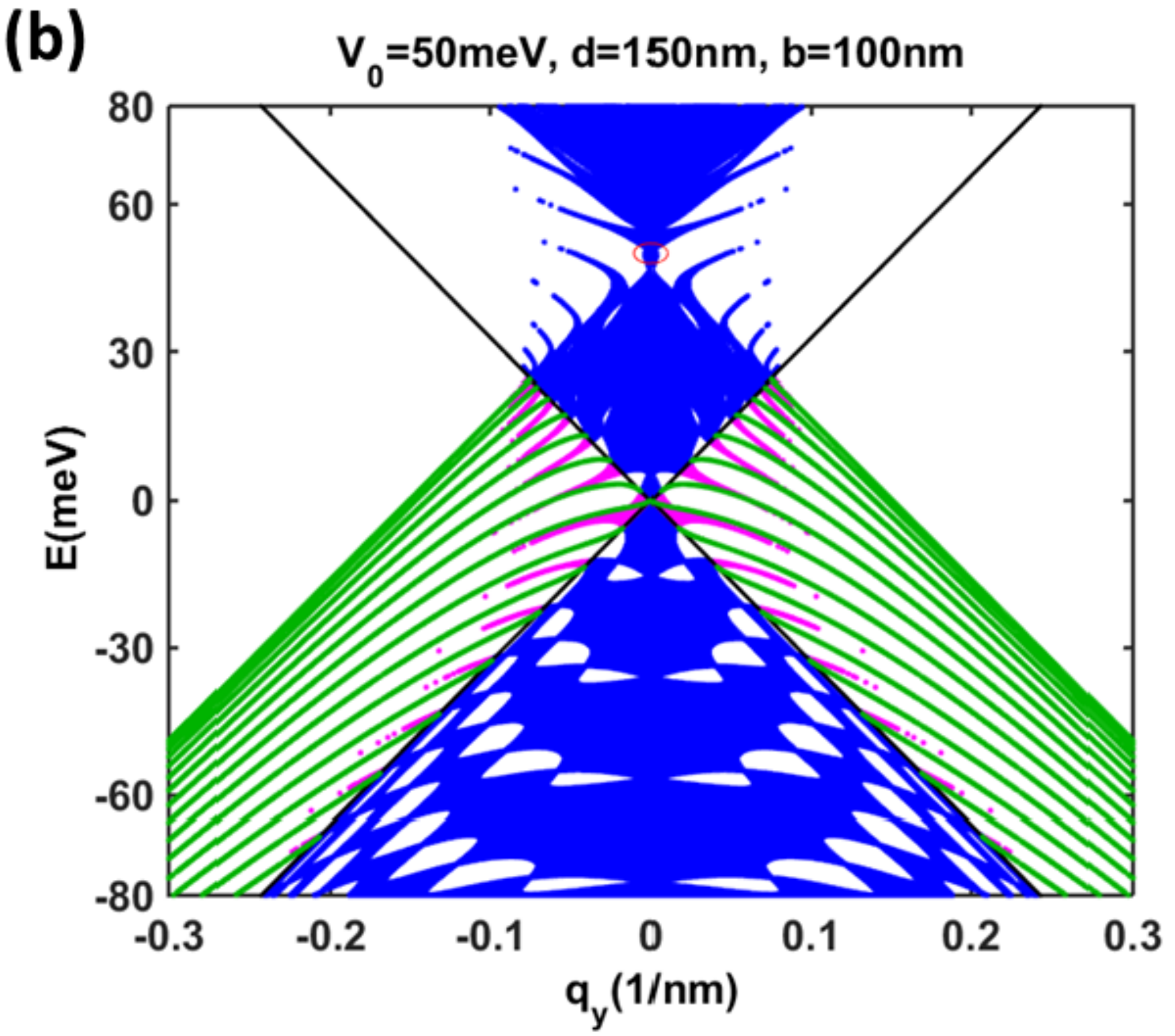} } \\
\subfloat{\includegraphics[width=0.5\columnwidth,height=0.4
\columnwidth]{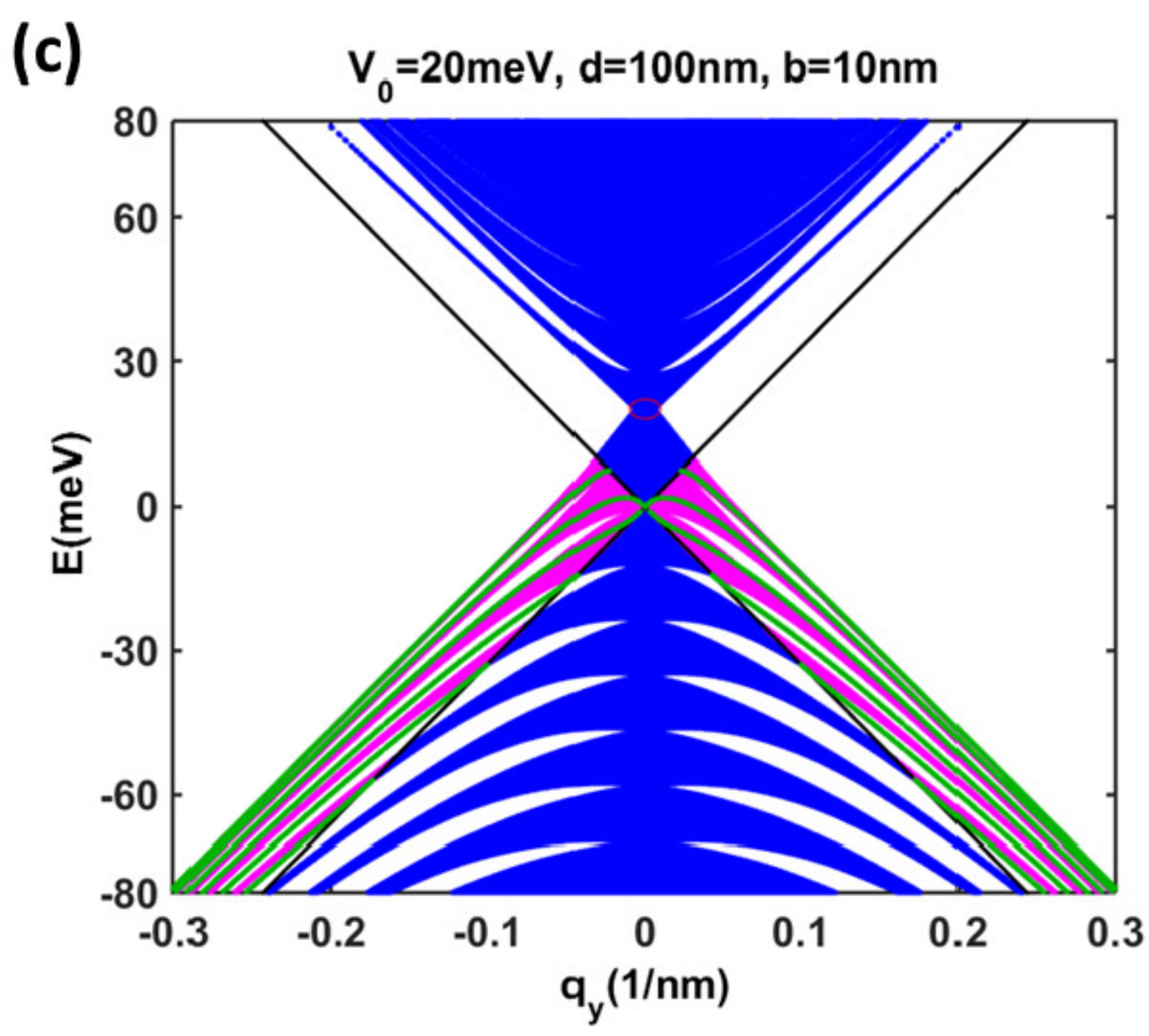} 
\label{fig:diracpoints-c} } 
\subfloat{\includegraphics[width=0.5\columnwidth,height=0.4
\columnwidth]{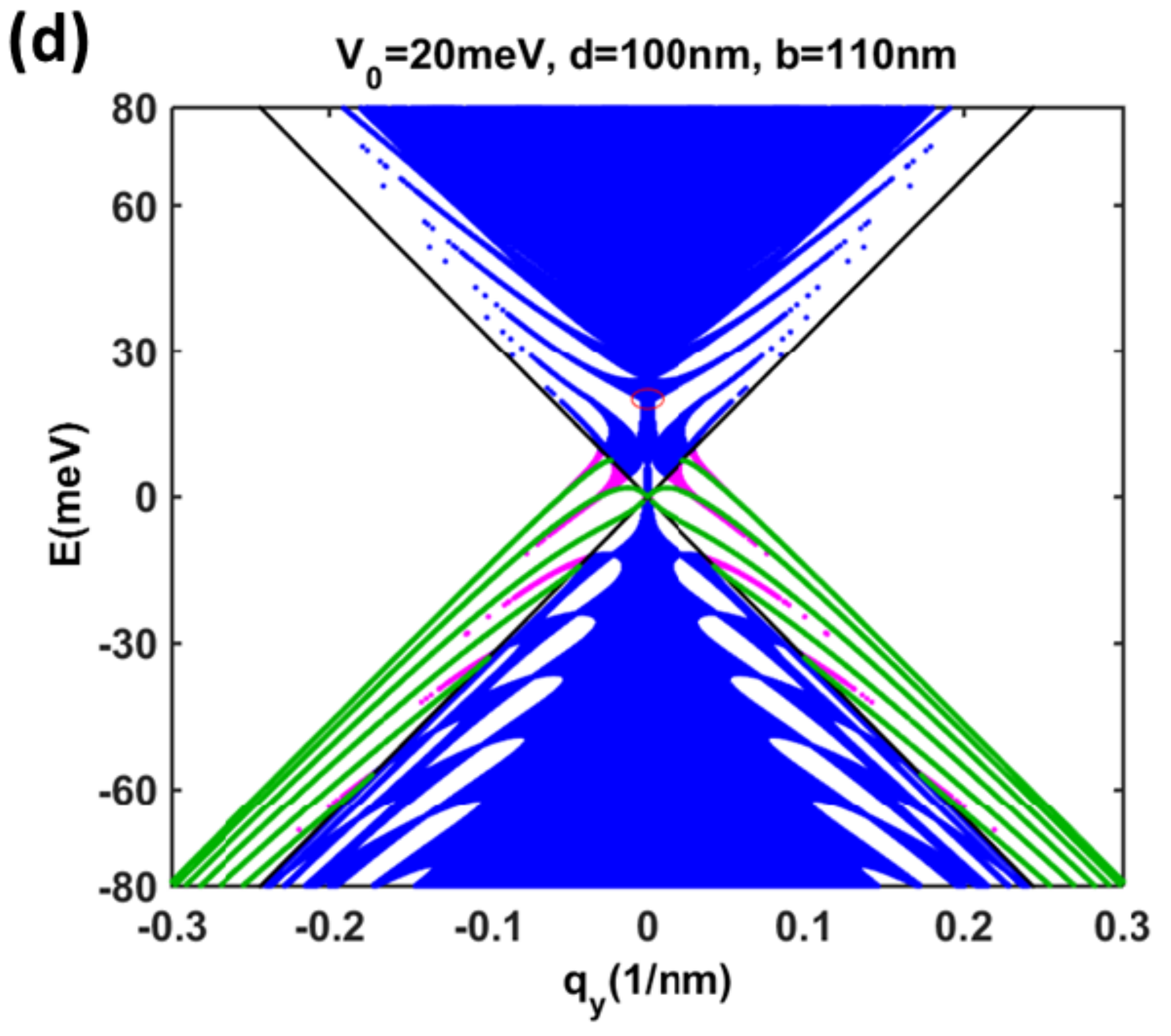} 
\label{fig:diracpoints-d} }
\caption{{\it (color online) }(a)-(d) Band structure for surface states of 3DTI for different potential barrier strength $V_{0}$, different barrier width(d) and barrier separation(b). Bands from the continuum states inside the dirac cone are colored blue. Bands outside the dirac cone due to bound states are colored magenta. Circle enclose the E=V point. The bound states in a single potential barrier (green) are superimposed on the band structure.}
\label{fig:boundbands} 
\end{figure}

Even though local measurement such as local DOS \cite{Kapitulnik, Balatsky} can detect such bound state formation by single barrier, creating a global effect on macroscopic sample will be more desirable for application.
An obvious way to achieve this is to tile the surface with a periodic array of such gate voltage induced potential barriers. Such tunable superlattice  structure of periodic potential have recently been realized for massless dirac fermions in the case of Graphene \cite{Marcus, Mandar} 
and recently in the case of Topological insulators \cite{Madhavan}. 
Very recently persistent optical gating of Topological Insualtor is acheived through which such gated structure \cite{Yeats} can also be acheived. 
We consider a periodic array of the potential barrier (see Fig.\ref{fig:periodic}(b)) given as 
\beq V(x) = \sum_{n} V_{0} \Theta( x + \frac{d}{2} - nL)\Theta ( nL + \frac{d}{2} -x) , n \in I   \label{periodic} \eeq
where $\Theta(x)$ is the Heaviside step function. Here $L$ is unit cell size, $b=L-d$ is the inter-barrier separation. 
In the $n-1$ th unit cell the wave functions in the region $I$ and $II$ are respectively given by 
\begin{widetext}
\bea
\psi_{I}(x) & = & C_{n-1}e^{iq_{x}(x-(n-1)L)}\left( \begin{array}{cc}
  1\\
  \pm e^{i\theta}
\end{array} \right)+D_{n-1} e^{-iq_{x}(x-(n-1)L)}\left( \begin{array}{cc}
  1\\
 \mp e^{-i\theta}
\end{array} \right)
\nonumber \\
\psi_{II} (x) & = & A_{n-1}e^{ik_{x}(x-(n-1)L)}\left( \begin{array}{cc}
  1\\
  \pm e^{i\phi}
\end{array} \right)+B_{n-1} e^{-ik_{x}(x-(n-1)L)}\left( \begin{array}{cc}
  1\\
 \mp e^{-i\phi}
\end{array} \right)
\eea 
\end{widetext}
The wave function in the $n$-th cell is given by 
 \beq
\psi(x) = C_{n}e^{iq_{x}(x-nL)}\left( \begin{array}{cc}
  1\\
  \pm e^{i\theta}
\end{array} \right)+D_{n} e^{-iq_{x}(x-nL)}\left( \begin{array}{cc}
  1\\
 \mp e^{-i\theta}
\end{array} \right)
\eeq
Matching the boundary conditions in the interfaces $x=L(n-1)-b$ and $x=(n-1)L$ we get ( for details of the method see \cite{Yeh, SGMS})
$ \begin{pmatrix}
C_{n-1}\\
D_{n-1}
\end{pmatrix} = M\begin{pmatrix}
C_{n}\\
D_{n}
\end{pmatrix}$
where $M=\begin{pmatrix}
M_{11} & M_{12}\\
M_{21} & M_{22}
\end{pmatrix}$
is unimodular transfer matrix that  connects equivalent unit cell.  However in a periodic potential Bloch equation demands  
\bea
\begin{pmatrix}
C_{n}\\
D_{n}
\end{pmatrix}=e^{iKL}\begin{pmatrix}
C_{n-1}\\
D_{n-1}
\end{pmatrix} .\nonumber
\eea
Equating these two relations one gets the eigenvalue condition as det$\vert M_{total}-\lambda I\vert =0$
whose solution gives the Bloch vector as 
\beq \lambda_{1}+\lambda_{2}=e^{-iKL}+e^{iKL} \Rightarrow 
K = \frac{1}{L} \cos^{-1} [ \frac{1}{2} Tr(M_{ij}) ] \label{Bloch} \eeq	
Eq. (\ref{Bloch}) when explicitly written in terms of the matrix element takes the usual Kronig-Penny form ( $k_{x} = i\kappa$) 
\begin{widetext} 
\bea
\cos KL & = & cos(k_{x}b)~cos(q_{x}d)+sin(k_{x}b)~sin(q_{x}d)[tan\theta ~ tan\phi -\dfrac{1}{cos\theta cos\phi}]~ \varepsilon > \bq
\label{KP1}  \\
\cos KL &=& cosh(\kappa b)~cos(q_{x}d)+sinh(\kappa b)~sin(q_{x}d)[tan\theta ~ coth\alpha -\dfrac{1}{cos\theta sinh\alpha}], 
\varepsilon < \bq  \label{KP2}
 \eea \end{widetext}
The band structures corresponding to Eq. (\ref{KP1}) and Eq. (\ref{KP2}) belongs two distinct region of bands in the $E-q_{y}$ 
plane, one within the dirac cone due to the presence of scattering states ( blue in Fig.\ref{fig:boundbands})
and the other region outside the dirac due to the bound states (magenta in Fig.\ref{fig:boundbands}), separated by 
$\varepsilon=\bq$. Such band structure is unique 
to the MDF because of the formation of bound states in a potential barrier and constitutes 
one of the most important results in this paper. The band structure that is formed within the dirac cone can again be analyzed to extract information
for a number of interesting properties such as additional dirac points \cite{Park, Brey, Spiral}, miniband formation \cite{Shen, Arun} which was already studied for dirac fermions in other contexts.

Here we only explain bands in the region $\varepsilon < \bq$ using the tight-binding approximation. 
These bands arise in a similar way like in a generic tight binding model due to the lifting of degeneracy of the bound states formed in each barrier by hopping amplitude. 
A typical example is that of the Landau bands in 
 Hofstadter butterfly \cite{Hofstadter} where the degeneracy of Landau levels are lifted by the introduction of a lattice potential 
However now they co-exist with the bands formed out of scattering states within the dirac cone which set them apart from the Hofstadter problem. 
When the barrier separation ($b$) relative to the barrier width ($d$) is increased, the hopping amplitude is decreased. This leads to the shrink of the band width and can be clearly seen by comparing the band structure in the left and right column of the Fig. \ref{fig:boundbands}. Since the number of bound states and their position changes with $V_{0}$, so do the band properties such as band gap, band position etc. Because of the discreteness of the resulting band structure over a wide range of barrier strength, it is expected that 
the DOS in presence of such periodic potential  will oscillate in a similar manner like $\rho_{b}$ in Eq. (\ref{DOSB}). This will in turn effect various properties of a system. 

\begin{figure}
\includegraphics[width=0.8\columnwidth , height= 
0.5\columnwidth]{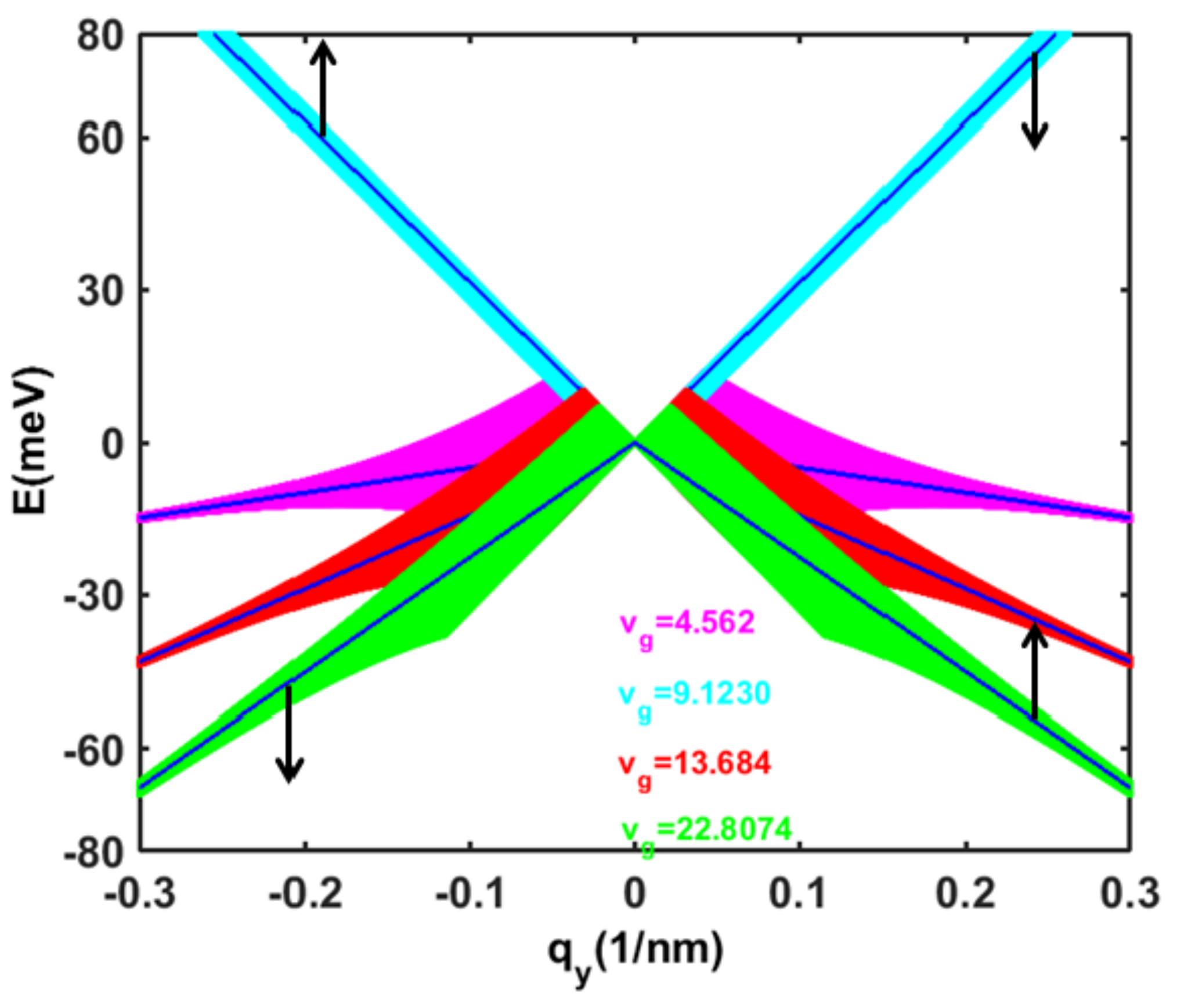} 
\caption{(Color online) The formation of bands by one dimensional helical modes for four different values of $Z$ or $v_{g}$. 
with corresponding dispersion of the helical states (lines) for a single $\delta$-function barrier superposed. 
Color of legends same as the color of bands. For two $v_{g}$'s the direction for spin for the helical states are indicated}
\label{fig:helical}
\end{figure}

It may be noted that in the band structure depicted in Fig. \ref{fig:boundbands}
a special situation arised when  $E=V$ within the dirac cone. This is because at that particular point the solution of dirac equation is different. 
Such point represents zero modes solution  which has been discussed in number of works \cite{Brey,Park} earlier. Briefly, at $E=V$, the dirac equation have the form
\beq
\left[\partial_{x}^{2}-k_{y}^{2} \right]\psi_{1,2}(x)=0 
\label{EVeq}
\eeq  The solution of Eq. (\ref{EVeq}) will have the form
\beq
\psi(x)  =  
  Ce^{k_{y}x}\left( \begin{array}{cc}
  1\\
   0 
\end{array} \right)+ De^{-k_{y}x}\left( \begin{array}{cc}
  0\\
   1 
\end{array} \right),~~\vert x \vert < d/2
\eeq
We have obtained transcendental equation for $E=V$ point by using the same transfer matrix method for scattering states solution within the dirac cone, namely 
\bea
cosKL &=& cos(k_{x}b)~cosh(k_{y}d) \nonumber \\
&  & \mbox{} +tan\phi ~sin(k_{x}b)sinh(k_{y}d),\varepsilon > \bq
\label{EV1} \eea 
To explore further the non-trivial effects due to the bound state formation we consider the limit $d \rightarrow 0$ and $V_{0} \rightarrow \infty$ such that $Z=V_{0}d$ constant.
 Substitution of this in (\ref{qx}) gives 
$q_{x}d=\frac{Z}{\hbar v_{F}}=v_{g}$ such that 
 Eq. (\ref{bound}) gives $\tan(v_{g}) = -\frac{\kappa}{\epsilon}$. 
The  dispersion relation of the corresponding states are $\epsilon=\pm q_{y}~ cos(v_{g})$. 
Substitution of these results in Eq. (\ref{KP1}) and Eq. (\ref{KP2}) gives 
\begin{widetext}
\bea 
cosKb &=&cos(k_{x}b)~cos(v_{g})+sin(k_{x}b)~sin(v_{g})\left(\frac{\epsilon}{k_{x}} \right), \varepsilon < \bq \label{KPL1} \\
cosKb &=& cosh(\kappa b)~cos(v_{g})+sinh(\kappa b)~sin(v_{g})\left( \frac{\epsilon}{\kappa} \right), \varepsilon > \bq \label{KPL2} \eea 
\end{widetext}
The band structure generated by the bound states in this limit given by Eq.  (\ref{KPL2}) are plotted in Fig. \ref{fig:helical}. Each band corresponds to a given value of $Z$.
It is known \cite{Balatsky, Nagaosa} that in this limit the bound states corresponding to the  
helical edge modes on the surface of a 3DTI at the interface of each potential barrier forming Tomonoga-Luttinger states \cite{Luttinger}. 
For such states the momentum is locked with the spin whose  sign (up/down) is determined by the $Z$. 
By changing electrostatic potential $V_{0}$ and thereby $Z$ one can flip the spin of such helical states. The bands showed in Fig. \ref{fig:helical} by such helical modes 
can therefore play very important role in spintronics \cite{books}. 

To summarize we show that band structure of MDF in a periodic array of potential barrier is distinguished from conventional band structure due to the existence 
of bands formed out of  bound states that exist outside the dirac cone. They can create experimentally observable effect and in suitable limit may  lead to possibility of interesting application. 
We thank K. Sengupta and D. Kumar for helpful discussion. PM is supported by a UGC fellowship and SG is partially supported by a UGC grant under UGC-UKIERI thematic partnership. 

\end{document}